\documentclass[onecolumn,12pt]{IEEEtran} 
\usepackage{amssymb}
\usepackage{amsthm,color}
\usepackage{latexsym}
\usepackage{graphicx}
\usepackage{fancyhdr}
\usepackage{bbm}
\usepackage{amsmath}
\usepackage{latexsym,amssymb,amsxtra,mathrsfs,times}

\renewcommand{\vec}[1]{{\bf #1}}

\newtheorem{thm}{Theorem}

\newtheorem{lemma}[thm]{Lemma}

\theoremstyle{definition}

\newcommand{\tr}{{\rm Tr}}
\newcommand{\R}{\mathbb R}
\newcommand{\G}{\mathcal{G}(\vec{s})}

\newcommand{\pp}{\Omega_{N_a}^{(a)} \times \Omega_{N_b}^{(b)}}

\newcommand{\F}{{\rm GF}}

\newcommand{\I}{\mathbf I}
\newcommand{\J}{\mathbf J}

\newcommand{\C}{{\mathcal C}}
\newcommand{\cf}{{\mathscr C}}
\newcommand{\CC}{{\mathbb C}}

\newcommand{\D}{{\mathcal D}}

\newcommand{\Ea}{{\mathbb{E}\left(A_l(\vec{s}), \Omega_{N_a}^{(a)} \times \Omega_{N_b}^{(b)}\right)}}
\newcommand{\E}{{\mathbb{E}}}

\newcommand{\A}{{A_l(\vec{s})}}

\newcommand{\MP}{\mathrm{MP}}

\newcommand{\ud}{\mathrm{d}}

\usepackage{setspace}
\begin{document}




\title{On a question of Babadi and Tarokh II}

\author{Jing Xia$^1$ \thanks{1. Fred Hutchinson Cancer Research Center, 1100 Fairview Ave N, Seattle, WA, USA}, Liuquan Wang$^2$ \thanks{2. Department of Mathematics, Zhejiang University, P.R. China}, Maosheng Xiong$^3$ \thanks{3. Department of Mathematics, Hong Kong University of Science and Technology, Clear Water Bay, Kowloon, Hong Kong}}





\maketitle


\renewcommand{\thefootnote}{}





\begin{abstract}

In this paper we continue to study a question proposed by Babadi and Tarokh \cite{ba2} on the mysterious randomness of Gold sequences. Upon improving their result, we establish the randomness of product of pseudorandom matrices formed from two linear block codes with respect to the empirical spectral distribution, if the dual distance of both codes is at least 5, hence providing an affirmative answer to the question.

\end{abstract}


\section{Introduction}\label{sec-into}

The elegant theory of random matrices (\cite{and,meh}), and in particular properties of the spectral distribution, have been studied for a long time but remain a prominent and active research area due to its wide and important applications in many diverse disciplines such as mathematical statistics, theoretical physics, number theory, and more recently in economics \cite{paf} and communication theory \cite{tul}. Most of the random models considered so far are matrices whose entries have i.i.d. structures. In a beautiful paper \cite{ba1}, Babadi and Tarokh considered matrices formed by choosing randomly codewords from a linear block code and proved the remarkable result that these matrices behave like random matrices of i.i.d. entries with respect to the so-called ``empirical spectral distribution'', if the dual distance of the code is sufficiently large. In a more recent work \cite{ba2}, investigating much further on the subject, Babadi and Torokh considered two matrices formed by choosing randomly codewords from two linear block codes and proved decisively that the products of such two matrices also behave like random matrices with respect to the empirical spectral distribution, if the dual distance of both codes is sufficiently large.

On the other hand, the authors (\cite{ba0,ba1,ba2}) already have observed by numerical experiments that matrices formed from Gold codes \cite{G} seem to behave like random matrices with respect to the empirical spectral distribution, even though the dual distance is as small as 5. Hence arises the natural question as to whether or not the stringent requirement of large dual distance could be relaxed in order to explain the mysterious randomness of Gold sequences. For matrices formed from liner block codes as considered in \cite{ba1}, an affirmative answer was recently provided by the first and the third authors (\cite{xx}) by using some ideas from number theory. Using similar ideas, in this paper we will prove the randomness of products of matrices formed from Gold sequences, hence improving upon the result of \cite{ba2}.

In order to describe the main result, we first give some notation. For the sake of generality, let $\F(q)$ be a finite field of order $q$ and let $\psi: \F(q) \to \CC^*$ be the standard additive character given by
\[\psi(z)=\exp\left(\frac{2 \pi \sqrt{-1} \, \tr_{q/l}(z)}{l}\right),\]
here $l$ is a prime number, $q$ is a power of $l$, and $\tr_{q/l}$ denotes the trace mapping from $\F(q)$ to $\F(l)$. When $q=l=2$, then $\psi(z)=(-1)^z \in \{-1,1\}$ for $z \in \F(2)$. In general it is known that $\psi(z)$ is a complex $l$-th root of unity.

Let $\C$ be an $[n,k,d]$ linear block code of length $n$, dimension $k$ and minimum Hamming distance $d$ over $\F(q)$. The dual code of $\C$, denoted by $\C^{\bot}$, is an $[n,n-k,d^{\bot}]$ linear block code over $\F(q)$ such that all the codewords of $\C^{\bot}$ are orthogonal to those of $\C$ with the natural inner product defined over $\F(q)^n$. Let $\epsilon: \F(q)^n \to (\CC^*)^n$ be the component-wise mapping $\epsilon(v_i):=\psi(v_i)$, for $\vec{v}=(v_1,v_2,\ldots,v_n) \in \F(q)^n$. For $p<n$, let ${\bf \Phi}_{\C}$ be a $p \times n$ random matrix whose rows are obtained by mapping a uniformly drawn set of size $p$ of the codewords of $\C$ under $\epsilon$. The \emph{Gram matrix} of the $p \times n$ matrix ${\bf \Phi}_{\C}$ is defined as $\mathcal{G}_{\C}:={\bf \Phi}_{\C}{\bf \Phi}_{\C}^*$, where ${\bf \Phi}_{\C}^*$ is the conjugate transpose of ${\bf \Phi}_{\C}$. Let $\{\lambda_1,\lambda_2,\ldots,\lambda_n\}$ be the set of eigenvalues of an $n \times n$ matrix $\vec{A}$. The \emph{spectral measure} of $\vec{A}$ is defined by
\[\mu_{\vec{A}}:=\frac{1}{n} \sum_{i=1}^n \delta_{\lambda_i}, \]
where $\delta_z$ is the Dirac measure. The empirical spectral distribution of $\vec{A}$ is defined as
\[M_{\vec{A}}(z):=\int_{-\infty}^z \mu_{\vec{A}}(\ud z). \]

The main result of this paper is as follows.

\begin{thm} \label{1:thm} Let $\C^a,\C^b$ be two linear block codes over $\F(q)$ of length $n$. Let $d_a^{\bot},d_b^{\bot}$ be the dual distances of $\C^a,\C^b$ respectively. Let $N_a,N_b$ be positive integers and $y_a=n/N_a, y_b=n/N_b$. Let $\vec{A}$ and $\vec{B}$ be two random matrices of size $N_a \times n$, $N_b \times n$ which are based on $\C_a$ and $\C_b$ respectively. Let $M_{\cf}(x)$ be the empirical spectral distribution function of the Gram matrix of $\frac{1}{\sqrt{N_aN_b}}\vec{A}\vec{B}^*$ and let $\overline{M}_{\boxplus^2\MP}(x;y_a,y_b)$ be the distribution of the free multiplicative convolution of the Marchenko-Pastur densities $\mu_{\MP}(x;y_a)$ and $\mu_{\MP}(x;y_b)$. Let
\[M_{\boxplus^2\MP}(x;y_a,y_b):=(1-y_a)+y_a\overline{M}_{\boxplus^2\MP}(x;y_a,y_b).\]
Suppose that $d^{\bot}:=\min\{d_a^{\bot},d_b^{\bot}\} \ge 5$ and $y_a,y_b \in (0,1) \cup (1,\infty)$. Then there are constants $C_1(y_a,y_b)$ and $C_2(y_a,y_b)$ depending only on $y_a,y_b$ such that for any $n \ge C_2(y_a,y_b)$ we have
\begin{eqnarray} \label{1:eqnthm} \sup_{x \in \R}\left|M_{\cf}(x)-M_{\boxplus^2\MP}(x;y_a,y_b)\right| \le \frac{C_1(y_a,y_b) \, \log \log n}{\log n} \,\,. \end{eqnarray}
\end{thm}

Interested readers may compare Theorem \ref{1:thm} with \cite[Theorems 2.3 and 2.4]{ba2}. It follows that the empirical spectral distribution of the Gram matrix of the random matrix $\frac{1}{\sqrt{N_aN_b}}\vec{A}\vec{B}^*$, with $\vec{A}$ and $\vec{B}$ based on linear block codes $\C^a$ and $\C^b$ respectively, resembles the universal empirical spectral distribution $M_{\boxplus^2\MP}(x;y_a,y_b)$ in the sense of Kolmogorov distance as $n \to \infty$, provided that the dual distances of $\C^a$ and $\C^b$ are both at least $5$. This provides an affirmative answer to the question related to the randomness of Gold sequences which was raised in \cite{ba2}. Moreover, as in \cite{xx}, the condition $d^{\bot} \ge 5$ in Theorem \ref{1:thm} could be slightly relaxed by assuming that the number of weight 4 codewords in $C^{\bot}$ is relatively small. On the other hand, if $d^{\bot}=3$, similar to \cite[Theorem 1]{xx}, it is quite unlikely that Theorem \ref{1:thm} remains true as was shown by Babadi, Ghassemzadeh and Tarokh (\cite[Theorem 3.1]{ba0}) on the remarkable example of shortened first-order Reed-Muller (Simplex) codes which have dual distance 3. Finally, it may be elementary to evaluate explicitly the constants $C_1(y_a,y_b),C_2(y_a,y_b)$, however, the process is very complicated, hence we choose not to do in this paper.

The proof of Theorem \ref{1:thm} follows the same strategy used in \cite{xx}, where some essence of number theory plays a prominent role in the study. We shall prove Theorem \ref{2:moment}, which improves \cite[Lemma 3.4]{ba2} substantially. Equipped with Theorem \ref{2:moment}, in Section \ref{sec-thm1} we will prove Theorem \ref{1:thm} directly following the argument of Babadi and Torokh (\cite{ba1,ba2}). 

\section{Estimate of the $l$-th moment}\label{sec-moment}

In this section we study the $l$-th moment of the empirical spectral distribution, similar to \cite[Lemma 3.4]{ba2}. We use slightly different notation.

As in Introduction, for $i=a,b$, let $\C^i$ be a linear block code over $\F(q)$ of length $n$ with dimension $k_i \ge 5$, and let $d_i^{\bot}$ be the dual distance of $\C^i$. Let $\epsilon: \F(q)^n \to (\CC^*)^n$ be the component-wise mapping. Define $\D^i=\epsilon(\C^i)$. For positive integers $N_i$, in order to choose randomly $N_i$ elements from $\D^i$, we define $\Omega^{(i)}_{N_i}$ to be the set of all maps $s: [1,N_i] \to \D^i$ endowed with the uniform probability, here $[1,N_i]$ denotes the set of integers from $1$ to $N_i$. Hence $\Omega^{(a)}_{N_a} \times \Omega^{(b)}_{N_b}$ is a probability space with cardinality $q^{k_aN_a+k_bN_b}$. For each $s^{(i)} \in \Omega_{N_i}^{(i)}$, the $N_i \times n$ matrix $\vec{A}_{s^{(i)}}$ corresponding to $s^{(i)}$ is given by
\[\vec{A}_{s^{(i)}}^T=\left[s^{(i)}(1)^T,s^{(i)}(2)^T, \ldots,s^{(i)}(N_i)^T \right]_{n \times N_i}\,\,, i=a,b,\]
here we have written $s^{(i)}(k) \in \D^{(i)}$ as a row vector. For any $\vec{u}=(u_1,\ldots,u_n),\vec{v}=(v_1,\ldots,v_n) \in \CC^n$, the (Hermitian) inner product is
\[\langle \vec{u},\vec{v} \rangle:=u_1 \bar{v}_1+\cdots+u_n \bar{v}_n.\]
Let $\mathcal{G}(\vec{s})$ be the Gram matrix of $\frac{1}{\sqrt{N_aN_b}} \vec{A}_{s^{(a)}}\vec{B}_{s^{(b)}}^*$. This is an $N_a \times N_a$ Hermitian matrix. Let $\lambda_1(\vec{s}), \ldots, \lambda_{N_a}(\vec{s})$ be the eigenvalues of $\mathcal{G}(\vec{s})$. For any positive integer $l$, define
\[\A:=\frac{1}{N_a} \sum_{i=1}^{N_a} \lambda_i(\vec{s})^l=\frac{1}{N_a}\,\, \tr \left(\G^l\right). \]
The purpose of this section is to compute $\Ea$, the $l$-th moment of the spectral measure. We prove a general result:

\begin{thm} \label{2:moment}
For $i=a,b$, let $y_i:=n/N_i$ and $Y_i:=\max\{1,y_i\}$. Let $d^{\bot}:=\min\{d_a^{\bot},d_b^{\bot}\}$. Assume that $d^{\bot} \ge 5$. Then for any $2 \le l <\min\{\sqrt{N_a},\sqrt{N_b}\}$, we have
\begin{eqnarray} \label{2:momenteqn}
\Ea=\sum_{i=1}^{l}y_a^{l-i+1} \sum_{\substack{k_1+k_2+\cdots+k_s=l-i+1\\
k_1+2k_2+\cdots+ik_i=l}}\frac{l!}{i!} \prod_{j=1}^i \frac{m_{\MP}^{(j)}(y_b)^{k_j}}{k_j!}+E_l,
\end{eqnarray}
where $m_{\MP}^{(l)}(y)$ is the $l$-th moment of the Marchenko-Pastur distribution $\mu_{\MP}$, given by
\[m_{\MP}^{(l)}(y):=\sum_{i=0}^{l-1} \frac{y^i}{i+1} \binom{l}{i} \binom{l-1}{i}, \]
and $E_l$ is bounded by
\[|E_l| \le \frac{l^{6l}\, Y_a(Y_aY_b)^l }{\min\{N_a,N_b\}} \,\,, \]
\end{thm}

Theorem \ref{2:moment} improves upon \cite[Lemma 3.4]{ba2} substantially. We remark that the main term on the right hand side of (\ref{2:momenteqn}) is off by a factor $y_a$, compared with \cite[Lemma 3.4]{ba2}. However, checking their proof carefully and also checking the paper \cite{bai}, it seems our formulation is correct. The rest of this section is devoted to a proof of Theorem \ref{2:moment}.

\subsection{Problem setting up}

We have
\[A_l(\vec{s})=\frac{1}{N_a^{l+1}N_b^l} \tr \left(\left(\vec{A}_{s^{(a)}}\vec{B}_{s^{(b)}}^*
\vec{B}_{s^{(b)}}\vec{A}_{s^{(a)}}^*\right)^l\right)=
\frac{1}{N_a^{l+1}N_b^l} \tr \left(\left(\vec{A}_{s^{(a)}}^*\vec{A}_{s^{(a)}}\vec{B}_{s^{(b)}}^*
\vec{B}_{s^{(b)}}\right)^l\right). \]
Noting that
\[\vec{A}_{s^{(a)}}^*\vec{A}_{s^{(a)}}=\sum_{i=1}^{N_a}s^{(a)}(i)^*s^{(a)}(i), \]
this gives
\begin{eqnarray} \label{2:tr1} \tr \left(\left(\vec{A}_{s^{(a)}}^*\vec{A}_{s^{(a)}}\vec{B}_{s^{(b)}}^*
\vec{B}_{s^{(b)}}\right)^l\right)=\sum_{\substack{1 \le i_1,\ldots,i_l \le N_a\\
1 \le j_1,\ldots,j_l \le N_b}}\tr\left(\prod_{k=1}^ls^{(a)}(i_k)^*s^{(a)}(i_k)s^{(b)}(j_k)^*s^{(b)}(j_k) \right).  \end{eqnarray}
The right hand is also
\[\sum_{\substack{1 \le i_1,\ldots,i_l \le N_a\\
1 \le j_1,\ldots,j_l \le N_b}}\tr\left(\prod_{k=1}^ls^{(a)}(i_k)s^{(b)}(j_k)^*s^{(b)}(j_k)s^{(a)}(i_{k+1})^* \right),\]
where the subscript index is modulo $l$, i.e., we use $i_{l+1}:=i_1$.

Both $s^{(a)}(i_k)s^{(b)}(j_k)^*$ and $s^{(b)}(j_k)s^{(a)}(i_{k+1})^*$ are real numbers. For $i=a,b$, denote by $\Pi_{N_i}^l$ the set of maps $\gamma: [0,l] \to [1,N_i]$. We may rewrite (\ref{2:tr1}) as
\begin{eqnarray*}  \tr \left(\left(\vec{A}_{s^{(a)}}^*\vec{A}_{s^{(a)}}\vec{B}_{s^{(b)}}^*
\vec{B}_{s^{(b)}}\right)^l\right)=\sum_{\substack{\gamma_a \in \Pi_{N_a}^l\\
\gamma_b \in \Pi_{N_b}^l}}\omega_{\underline{\gamma}}(\vec{s}),   \end{eqnarray*}
where
\[\omega_{\underline{\gamma}}(\vec{s}):=\prod_{k=1}^l \left\{s^{(a)}(\gamma_a(k)) s^{(b)}(\gamma_b(k))^* s^{(b)}(\gamma_b(k)) s^{(a)}(\gamma_a(k+1))^*\right\}.\]
Here again we have used modulo $l$ if necessary. Hence we have
\[A_l:=\E\left(A_l(\vec{s}),\Omega^{(a)}_{N_a} \times \Omega^{(b)}_{N_b}\right)=\frac{1}{N_a^{l+1}N_b^l}\sum_{\substack{\gamma_a \in \Pi_{N_a}^l\\
\gamma_b \in \Pi_{N_b}^l}}\E\left(\omega_{\underline{\gamma}}(\vec{s}),\Omega^{(a)}_{N_a} \times \Omega^{(b)}_{N_b}\right). \]

For $i=a,b$, let $\Sigma_{N_i}$ be the group of permutations of the set $[1,N_i]$. Then $\Sigma_{N_i}$ acts on $\Pi_{N_i}^l$, since $\sigma \circ \gamma_i \in \Pi_{N_i}^l$ whenever $\gamma_i \in \Pi_{N_i}^l$ and $\sigma \in \Sigma_{N_i}$. Let $[\gamma_i]$ be the equivalent class of $\gamma_i$, that is,
\[[\gamma_i]=\{\sigma \circ \gamma_i: \sigma \in \Sigma_{N_i} \}. \]
We may write
\[A_l=\frac{1}{N_a^{l+1}N_b^l}\sum_{\substack{\gamma_i \in \Pi_{N_i}^l/\Sigma_{N_i}\\
i=a,b}} \,\,\, \sum_{\substack{\tau_i \in [\gamma_i]\\
i=a,b}}\E\left(\omega_{\underline{\tau}}(\vec{s}),\Omega^{(a)}_{N_a} \times \Omega^{(b)}_{N_b}\right). \]
For any fixed $\sigma_i \in \Sigma_{N_i}$, as $s^{(i)}$ runs over $\Omega_{N_i}^{(i)}$, clearly $s^{(i)} \circ \sigma_i$ also runs over $\Omega_{N_i}^{(i)}$, hence
\[\E\left(\omega_{\underline{\sigma_i \circ \gamma_i}} (\vec{s}),\Omega^{(a)}_{N_a} \times \Omega^{(b)}_{N_b}\right)=\E\left(\omega_{\underline{\gamma}}(\underline{s^{(i)} \circ \sigma_i}),\Omega^{(a)}_{N_a} \times \Omega^{(b)}_{N_b}\right)=\E\left(\omega_{\underline{\gamma}}(\vec{s}),\Omega^{(a)}_{N_a} \times \Omega^{(b)}_{N_b}\right). \]
Moreover, for $i=a,b$, let
\[V_{\gamma_i}:=\gamma_i \left([0,l]\right) , \quad v_{\gamma_i}:=\# V_{\gamma_i},\]
and define the probability space
\[\Omega(V_{\gamma_i}):=\{s^{(i)}: V_{\gamma_i} \to \D^i\}\]
assigned with the uniform probability. It is clear that $\#[\gamma_i]=\frac{N_i!}{(N_i-v_{\gamma_i})!}, \#\Omega(V_{\gamma_i})=q^{k_iv_{\gamma_i}}$ and
\[\E\left(\omega_{\underline{\gamma}}(\vec{s}),\Omega^{(a)}_{N_a} \times \Omega^{(b)}_{N_b}\right)=\E\left(\omega_{\underline{\gamma}}(\vec{s}),\Omega(V_{\gamma_a}) \times \Omega(V_{\gamma_b})\right). \]
Summarizing the above we have
\begin{eqnarray} \label{2:set} A_l=\frac{1}{N_a^{l+1}N_b^l}\sum_{\substack{\gamma_i \in \Pi_{N_i}^l/\Sigma_{N_i}\\
i=a,b}} \frac{N_a! \, N_b!}{(N_a-v_{\gamma_a})! \, (N_b-v_{\gamma_b})!} \,\, \E\left(\omega_{\underline{\gamma}}(\vec{s}),\Omega(V_{\gamma_a}) \times \Omega(V_{\gamma_b})\right). \end{eqnarray}

\subsection{Study of $\E\left(\omega_{\underline{\gamma}}(\vec{s}),\Omega(V_{\gamma_a}) \times \Omega(V_{\gamma_b})\right)$}

For simplicity, we may write $\E\left(\omega_{\underline{\gamma}}(\vec{s}),\Omega(V_{\gamma_a}) \times \Omega(V_{\gamma_b})\right)$ as $W_{\underline{\gamma}}$. Suppose
\begin{eqnarray} \label{3:ia} V_{\gamma_a}&=&\{z_{\lambda}: 1 \le \lambda \le v_{\gamma_a}\}, \quad \I_{\lambda}:= (\gamma_a)^{-1}(z_{\lambda}),\\
\label{3:jb} V_{\gamma_b} &=& \{y_{\mu}: 1 \le \mu \le v_{\gamma_b}\}, \quad \J_{\mu}:= (\gamma_b)^{-1}(y_{\mu}). \end{eqnarray}
We define
\[\I_{\lambda,\mu}:=\I_{\lambda} \cap \J_{\mu}, \quad \left|\I_{\lambda,\mu}\right|=\delta(\lambda,\mu), \]
\[\widetilde{\I}_{\lambda,\mu}:=\widetilde{\I}_{\lambda} \cap \J_{\mu}, \quad \left|\I_{\lambda,\mu}\right|=\widetilde{\delta}(\lambda,\mu), \]
where $\widetilde{\I}_{\lambda}:=\I_{\lambda}-1$, i.e., $u \in \widetilde{\I}_{\lambda}$ if and only if $u+1 \pmod{l} \in \I_{\lambda}$. Now we have
\[W_{\underline{\gamma}}= q^{-k_av_{\gamma_a}-k_bv_{\gamma_b}} \sum_{\substack{s_i \in \Omega(V_{\gamma_i})\\
i=a,b}} \prod_{(\lambda,\mu)} \left\{s_a(z_{\lambda})s_b(y_{\mu})^*\right\}^{\delta(\lambda,\mu)}
\left\{s_b(y_{\mu})s_a(z_{\lambda})^*\right\}^{\widetilde{\delta}(\lambda,\mu)}.\]

For $i=a,b$, let
\[ H^{(i)T}=\left[\vec{h}_1^{(i)T},\vec{h}_2^{(i)T},
\ldots,\vec{h}_n^{(i)T}\right]\]
be a generating matrix of $\C^i$, where $\vec{h}_t^{(i)}=\left[h^{(i)}_{t1},h^{(i)}_{t2},\ldots,h^{(i)}_{tk_i}\right]$ is the $t$-th row vector. So each codeword of $\C^i$ is given by
\begin{eqnarray} \label{2:codec} c^i(\vec{x})=H^{(i)} [x_1,\ldots,x_{k_i}]^T,  \end{eqnarray}
for some unique $\vec{x}=(x_1,\ldots,x_{k_i}) \in \F(q)^{k_i}$. Hence each $s^{(i)}(u) \in \D^i$ corresponds to a unique length $k_i$ column-vector, which we may record as $\overrightarrow{s^{(i)}(u)} \in \F(q)^{k_i}$. From (\ref{2:codec}), the $t$-th entry of $s^{(i)}(u)$ is given by
\[s^{(i)}(u)[t]=\psi\left(\vec{h}_{t}^{(i)} \cdot \overrightarrow{s^{(i)}(u)}\right), \]
where $\psi:\F(q) \to \CC^*$ is the standard additive character. So
\[ s^{(a)} (z_{\lambda}) s^{(b)} (y_{\mu})^* = \sum_{t=1}^n \psi\left(\vec{h}_{t}^{(a)} \cdot \overrightarrow{s^{(a)}(z_{\lambda})}-\vec{h}_{t}^{(b)} \cdot \overrightarrow{s^{(b)}(y_{\mu})}\right), \]
and
\[ s^{(b)} (y_{\mu}) s^{(a)} (z_{\lambda})^* = \sum_{t=1}^n \psi\left(\vec{h}_{t}^{(b)} \cdot \overrightarrow{s^{(b)}(y_{\mu})}-\vec{h}_{t}^{(a)} \cdot \overrightarrow{s^{(a)}(z_{\lambda})}\right). \]
From this we find that
\begin{eqnarray*} W_{\underline{\gamma}} &=& \sum_{\substack{1 \le i_1^{(\lambda,\mu)},\ldots,i_{\delta(\lambda,\mu)}^{(\lambda,\mu)} \le n,\\
1 \le k_1^{(\lambda,\mu)},\ldots,k_{\widetilde{\delta}(\lambda,\mu)}^{(\lambda,\mu)} \le n,\\
1 \le \lambda \le v_{\gamma_a}\\
1 \le \mu \le v_{\gamma_b}}} \sum_{\substack{\overrightarrow{s^{(a)}(z_{\lambda})} \in \F(q)^{k_a}, \\
\overrightarrow{s^{(b)}(y_{\mu})} \in \F(q)^{k_b},\\
\forall \lambda, \mu }}
\psi \Bigg\{\sum_{(\lambda,\mu)} \left(\sum_{j=1}^{\delta(\lambda,\mu)} \vec{h}_{i_j^{(\lambda,\mu)}}^{(a)}- \sum_{j=1}^{\widetilde{\delta}(\lambda,\mu)} \vec{h}_{k_j^{(\lambda,\mu)}}^{(a)} \right) \overrightarrow{s^{(a)}(z_{\lambda})} \\
& &  -\sum_{(\lambda,\mu)} \left(\sum_{j=1}^{\delta(\lambda,\mu)} \vec{h}_{i_j^{(\lambda,\mu)}}^{(b)}- \sum_{j=1}^{\widetilde{\delta}(\lambda,\mu)} \vec{h}_{k_j^{(\lambda,\mu)}}^{(b)} \right) \overrightarrow{s^{(b)}(y_{\mu})} \Bigg\} \times q^{-k_av_{\gamma_a}-k_bv_{\gamma_b}} .
\end{eqnarray*}
Using the orthogonality property
\[\sum_{z \in \F(q)} \psi(zx)=\left\{\begin{array}{lll}
0&:& \mbox{ if } x \in \F(q) \setminus \{0\},\\
q&:& \mbox{ if } x=0,
\end{array}
\right.\]
we observe that we must have
\begin{eqnarray*}
\sum_{\mu=1}^{v_{\gamma_b}} \left(\sum_{j=1}^{\delta(\lambda,\mu)} \vec{h}_{i_j^{(\lambda,\mu)}}^{(a)}- \sum_{j=1}^{\widetilde{\delta}(\lambda,\mu)} \vec{h}_{k_j^{(\lambda,\mu)}}^{(a)} \right) & = & \vec{0}, \, \forall \, \lambda \\
\sum_{\lambda=1}^{v_{\gamma_a}} \left(\sum_{j=1}^{\delta(\lambda,\mu)} \vec{h}_{i_j^{(\lambda,\mu)}}^{(b)}- \sum_{j=1}^{\widetilde{\delta}(\lambda,\mu)} \vec{h}_{k_j^{(\lambda,\mu)}}^{(b)} \right) & = & \vec{0}, \, \forall \, \mu. \end{eqnarray*}
Otherwise the contribution on the right hand side to $W_{\underline{\gamma}}$ is zero. Writing in a different form, we conclude that the quantity $W_{\underline{\gamma}}$ is equal to the number of solutions $(t_1,t_2,\ldots,t_{l},\tau_1,\tau_2,\ldots,\tau_l)$ such that $1 \le t_1,t_2,\ldots,t_{l},\tau_1,\tau_2,\ldots,\tau_l \le n$ and the following two equations are satisfied:
\begin{eqnarray*} \sum_{u \in \I_{\lambda}} \vec{h}_{t_u}^{(a)} &=& \sum_{u \in \widetilde{\I}_{\lambda}} \vec{h}_{\tau_{u}}^{(a)} , \quad \forall \, \, 1 \le \lambda \le v_{\gamma_a}, \\
\sum_{u \in \J_{\mu}} \vec{h}_{t_u}^{(b)} &=& \sum_{u \in \J_{\mu}} \vec{h}_{\tau_{u}}^{(b)} , \quad \forall \, \, 1 \le \mu \le v_{\gamma_b},
\end{eqnarray*}
where $\I_{\lambda},\J_{\mu}$'s are given in (\ref{3:ia}) and (\ref{3:jb}).

\subsection{Study of $W_{\underline{\gamma}}$}

We first consider the system of linear equations over $\R$
\begin{eqnarray*} \sum_{u \in \I_{\lambda}} X_{u} &=& \sum_{u \in \widetilde{\I}_{\lambda}} Y_{u} , \quad \forall \, \, 1 \le \lambda \le v_{\gamma_a}, \\
\sum_{u \in \J_{\mu}} X_{u} &=& \sum_{u \in \J_{\mu}} Y_{u} , \quad \forall \, \, 1 \le \mu \le v_{\gamma_b},
\end{eqnarray*}
on the variables $X_1,\ldots,X_l,Y_1,\ldots,Y_l$. Let $W$ be the vector space of the set of solutions. We prove
\begin{lemma} \label{3:lem1} $\dim_{\R}W=2l-v_{\gamma_a}-v_{\gamma_b}+1$.
\end{lemma}

\noindent {\bf Proof.} It suffices to show that for any fixed real numbers $a_1,\ldots,a_{v_{\gamma_a}},b_1,\ldots,b_{\gamma_b}$, suppose that
\begin{eqnarray} \label{3:iden} \sum_{\lambda=1}^{v_{\gamma_a}}a_{\lambda} \left(\sum_{u \in \I_{\lambda}} X_{u} - \sum_{u \in \widetilde{\I}_{\lambda}} Y_{u} \right) - \sum_{\mu=1}^{v_{\gamma_b}}b_{\lambda} \left(\sum_{u \in \J_{\mu}} X_{u} - \sum_{u \in \J_{\mu}} Y_{u} \right) \equiv 0, \end{eqnarray}
then we must have
\[a_1=a_2=\cdots=a_{\gamma_a}=b_1=b_2=\cdots=b_{\gamma_b}. \]
Since (\ref{3:iden}) is an identity for any $X_u,Y_v$'s, the coefficients in front of any $X_u$ and $Y_v$ must be zero, hence we have
\begin{eqnarray*} a_{\lambda}-b_{\mu}=0 \quad \mbox{whenever} \quad \left(\I_{\lambda} \cup \widetilde{\I}_{\lambda} \right) \cap \J_{\mu} \ne \Phi. \end{eqnarray*}
Let $G$ be a bipartite graph with vertices $a_1,\ldots,a_{\gamma_a},b_1,\ldots,b_{\gamma_b}$ such that $a_{\lambda}$ and $b_{\mu}$ are connected whenever $\left(\I_{\lambda} \cup \widetilde{\I}_{\lambda} \right) \cap \J_{\mu} \ne \Phi$, and let $S$ be a maximal connected component of $G$, whose vertex set, without loss of generality, may be written as $S=\{a_1,\ldots,a_t,b_1,\ldots,b_s\}$. Then we have
\[a_1=\cdots=a_t=b_1=\cdots=b_s. \]
Define
\[A=\bigcup_{i=1}^t \left(\I_{i} \cup \widetilde{\I}_{i} \right), \quad B=\bigcup_{i=1}^s \J_{i}. \]
For any $u \in [1,l] \setminus B$, then $u \in \I_{\lambda}$ for some $\lambda$. Since $S$ is a maximal connected component, we must have $\lambda \notin \{1,\ldots,t\}$, and hence $u \in [1,l] \setminus A$. Therefore $A \subset B$. Similarly we have $B \subset A$. Thus we have
\[A=B.\]
Since $S$ is a maximal connected component, this implies that
\begin{eqnarray} \label{3:bipar}
\bigcup_{i=1}^t\left(\I_{i} \cup \widetilde{\I}_{i} \right) \cap
\bigcup_{j=t+1}^{\gamma_a}\left(\I_{j} \cup \widetilde{\I}_{j} \right) =\Phi. \end{eqnarray}
We prove from (\ref{3:bipar}) that
\[\bigcup_{i=1}^t\left(\I_{i} \cup \widetilde{\I}_{i} \right)=[1,l]. \]
This can be proved as follows: write
\[A_1:=\bigcup_{i=1}^t \I_{i} = \{a_1,a_2,\ldots,a_N\} \subset [1,l], \]
where $1 \le a_1 <a_2<\cdots<a_N \le l$. Then
\[B_1:=\bigcup_{i=1}^t \widetilde{\I}_{i} =\left\{a_1-1 \bmod{l},a_2-1,\ldots,a_N-1 \right\}. \]
If $a_1 \ge 2$, then $1 \le a_1-1 \notin A_1$, hence $a_1-1 \in \I_{\lambda}$ for some $\lambda \in \{t+1, \ldots,\gamma_a\}$, but we know $a_1-1 \in B_1$. So the requirement (\ref{3:bipar}) can not be met, contradiction. Hence we must have $a_1=1$.

We also have $a_2 \ge 2$. If $a_2 \ge 3$, then $2 \le a_2-1 \notin A_1$, by similar argument, we shall find a contradiction to (\ref{3:bipar}). Hence we have $a_2=2$.

Using this argument inductively, we shall find that $a_i=i$ for each $1 \le i \le N$. If $N<l$, then $l \notin A_1$. Noticing that $l \in B_1$, by using similar argument again we find contradiction. Hence $N=l$. We conclude that $A_1=[1,l]$. The completes the proof of Lemma \ref{3:lem1}. \quad $\square$

Now we assume that $d^{\bot}:=\min\{d_a^{\bot},d_b^{\bot}\} \ge 5$, that is, any four rows of $H^{(i)}, i=a,b$ are linearly independent. It follows from Lemma \ref{3:lem1} that $W_{\underline{\gamma}} \le n^{2l-v_{\gamma_a}-v_{\gamma_b}+1}$. Denote by $\Gamma$ the set of $\underline{\gamma}=(\gamma_a,\gamma_b)$ such that $W_{\underline{\gamma}} = n^{2l-v_{\gamma_a}-v_{\gamma_b}+1}$. We prove

\begin{lemma} \label{3:lem2} Assume that $d^{\bot}:=\min\{d_a^{\bot},d_b^{\bot}\} \ge 5$. Then
\[\left\{\begin{array}{lcc}
W_{\underline{\gamma}} = n^{2l-v_{\gamma_a}-v_{\gamma_b}+1}&:& \mbox{ if } \underline{\gamma} \in \Gamma, \\
W_{\underline{\gamma}} \le 4 n^{2l-v_{\gamma_a}-v_{\gamma_b}}&:& \mbox{ if } \underline{\gamma} \notin \Gamma.
\end{array} \right.
\]
\end{lemma}

\noindent {\bf Proof.} We first note that for $\underline{\gamma} \in \Gamma$, the equations in $W_{\underline{\gamma}}$ can be solved completely in the form of $t_u=\tau_v$ for some $u,v$'s, hence $W_{\underline{\gamma}} = n^{2l-v_{\gamma_a}-v_{\gamma_b}+1}$ from Lemma \ref{3:lem1}. If $\underline{\gamma} \notin \Gamma$, then we can not solve $W_{\underline{\gamma}}$ completely in this form, so there are two variable, say $t_1,t_2 \in \I_1$, such that
\[\vec{h}_{t_1}^{(a)}+\vec{h}_{t_2}^{(a)}+\cdots = \cdots + \cdots. \]
Given any values from $1$ to $n$ to all other variables, the number of different ways of doing that is $n^{2l-v_{\gamma_a}-v_{\gamma_b}}$ because of Lemma \ref{3:lem1}, we may need to solve the equation for $t_1,t_2$ such that
\[\vec{h}_{t_1}^{(a)}+\vec{h}_{t_2}^{(a)}=\vec{v} \]
for some $\vec{v}$ depending on all other variables except $t_1,t_2$. If $\vec{v}=\vec{0}$, this enforces a new relation on other variables, hence the number of ways such that $\vec{v}=\vec{0}$ is at most $n^{2l-v_{\gamma_a}-v_{\gamma_b}-1}$. On the other hand, for each given $t_1$, there is at most one value $t_2$ such that $\vec{h}_{t_1}+\vec{h}_{t_2}=\vec{0}$. Hence the total number of solutions for this case is at most $n^{2l-v_{\gamma_a}-v_{\gamma_b}}$. Let us define
\[A_{\vec{v}}=|\left\{(t_1,t_2): 1 \le t_1,t_2 \le n, \mbox{ and } \vec{h}_{t_1}+\vec{h}_{t_2}=\vec{v} \right\}|. \]
We have just proved that
\begin{eqnarray*}  W_{\underline{\gamma}} \le n^{2l-v_{\gamma_a}-v_{\gamma_b}} \left(1+\sup_{\vec{v} \ne \vec{0}}A_{\vec{v}}\right). \end{eqnarray*}
We have proved in \cite[Section IV]{xx} that if $d^{\bot} \ge 5$, then
\[A_{\vec{v}} \le 3, \quad \mbox{ if } \vec{v} \ne \vec{0}. \]
This implies that
\begin{eqnarray*} W_{\underline{\gamma}} \le 4 n^{2l-v_{\gamma_a}-v_{\gamma_b}} . \end{eqnarray*}
This completes the proof of Lemma \ref{3:lem2}. \quad $\square$

\subsection{Proof of Theorem \ref{2:moment}}

The equation (\ref{2:set}) can now be written as
\begin{eqnarray*} A_l=\frac{1}{N_a^{l+1}N_b^l}\sum_{\substack{\gamma_i \in \Pi_{N_i}^l/\Sigma_{N_i}\\
i=a,b}} \frac{N_a! \, N_b!}{(N_a-v_{\gamma_a})! \, (N_b-v_{\gamma_b})!} \,\, W_{\underline{\gamma}}. \end{eqnarray*}

Suppose $2 \le l <\sqrt{N_i}$ for $i=a,b$. Using $W_{\underline{\gamma}} \le 4n^{2l-v_{\gamma_a}-v_{\gamma_b}+1}$ and
\[N_i^{v_{\gamma_i}} \ge \frac{N_i!}{(N_i-v_{\gamma_i})!}>N_i^{v_{\gamma_i}} \left(1-v_{\gamma_i}/N_i\right)^{v_{\gamma_i}} \ge N_i^{v_{\gamma_i}} \left(1-v_{\gamma_i}^2/N_i\right), \]
we find
\begin{eqnarray} \label{3:al} A_l=\frac{1}{N_a^{l+1}N_b^l}\sum_{\substack{\gamma_i \in \Pi_{N_i}^l/\Sigma_{N_i}\\
i=a,b}} N_a^{\gamma_a} N_b^{\gamma_b} \, W_{\underline{\gamma}}+E_1,  \end{eqnarray}
where $E_1$ is bounded by
\[|E_1| \le 4\sum_{\substack{\gamma_i \in \Pi_{N_i}^l/\Sigma_{N_i}\\
i=a,b}} \left(\frac{n}{N_a}\right)^{l-v_{\gamma_a}+1} \left(\frac{n}{N_b}\right)^{l-v_{\gamma_b}} \frac{2l^2}{\min\{N_a,N_b\}}
\le \frac{8l^{2l+4}Y_a^{l+1}Y_b^l}{\min\{N_a,N_b\}}, \]
where $Y_i:=\max\{1,y_i\}$ and $y_i=n/N_i$ for $i=a,b$.

From Lemma \ref{3:lem2}, the contribution to $A_l$ from $\underline{\gamma} \notin \Gamma$ is bounded by
\[|E_2| \le \frac{4}{N_a^{l+1}N_b^l}\sum_{\substack{\gamma_i \in \Pi_{N_i}^l/\Sigma_{N_i}\\
i=a,b}} N_a^{v_{\gamma_a}} \, N_b^{v_{\gamma_b}}\, n^{2l-v_{\gamma_a}-v_{\gamma_b}}. \,\, . \]
It is easy to see that
\[|E_2| \le \frac{4 l^{2l}}{N_a}\sum_{1 \le u,v \le l} \left(\frac{n}{N_a}\right)^{l-u} \left(\frac{n}{N_b}\right)^{l-v} \le \frac{4 l^{2l+2} \left(Y_aY_b\right)^l}{N_a}\, , \]

On the other hand, it can be seen, from the combinatorial nature of $\Gamma$ and by consulting Lemmas 2.1 and 2.2 and the way of deriving equation (5.10) in \cite{bai}, that we shall find
\[\sum_{\substack{\underline{\gamma}=(\gamma_a,\gamma_b) \in \Gamma\\
\gamma_i \in \Pi_{l,N_i}/\Sigma_{N_i}, i=a,b \\
v_{\gamma_a}=u}} y_b^{l-v_{\gamma_b}}= \sum_{\substack{k_1+k_2+\cdots+k_u=l-u+1\\
k_1+2k_2+\cdots+uk_u=l}}\frac{l!}{u!} \prod_{j=1}^u \frac{m_{\MP}^{(j)}(y_b)^{k_j}}{k_j!}\,. \]
Returning to $A_l$ in (\ref{3:al}) where the main term comes from $\underline{\gamma}$'s such that $\underline{\gamma} \in \Gamma$ and combining all the above, we finish the proof of Theorem \ref{2:moment}. \quad $\square$

\section{Theorem \ref{1:thm}} \label{sec-thm1}

To prove Theorem \ref{1:thm}, we follow the method of \cite{ba1,ba2}. We need the following lemma from probability theory, which is discussed in details in \cite[Ch. XVI-3]{fel} (or see \cite[Lemma 3.1]{ba1}):

\begin{lemma} \label{3:prob} Let $F$ be a probability distribution with vanishing expectation and characteristic function $\phi$. Suppose that $F-G$ vanishes at $\pm \infty$ and that $G$ has a derivative $g$ such that $|g| \le m$. Finally, suppose that $g$ has a continuously differentiable Fourier transform $\gamma$ such that $\gamma(0)=1$ and $\gamma'(0)=0$. Then, for all $z$ and $T>0$ we have
\[|F(z)-G(z)| \le \frac{1}{\pi} \int_{-T}^T \left|\frac{\phi(t)-\gamma(t)}{t}\right|\, \ud t+\frac{24 m}{\pi T}. \]
\end{lemma}

\subsection{Some lemmas} Fix $y_b \in (0,1) \cup (1,\infty)$, the $l$-th moment of a Marchenko-Pastur distribution is given by
\begin{eqnarray} \label{3:mom} m_{\MP}^{(l)}=\sum_{i=0}^{l-1}\frac{y_b^i}{i+1} \binom{l}{i} \binom{l-1}{i}. \end{eqnarray}
We first prove

\begin{lemma} \label{3:lem1} For any $l \ge 1$ we have
\begin{eqnarray} \label{3:bmp} \left|m_{\MP}^{(l)}\right| < (8e^2)^{l}Y_b^l\, . \end{eqnarray}
\end{lemma}

\noindent {\bf Proof.} Elementary estimates on binomial coefficients yield
\[\left|m_{\MP}^{(l)}\right|<  \sum_{i=0}^{l-1} \frac{y_b^i l^{2i}}{(i!)^2} \le l \max_{0 \le i \le l-1} \frac{(y_bl^2)^i}{(i!)^2}. \]
By quotient test we find that the maximal value of $\frac{(yl^2)^i}{(i!)^2}$ is attained at $i=i_0=[\sqrt{y_b}l]$. If $y_b \ge 1$, then $\frac{(yl^2)^i}{(i!)^2}$ is increasing for $0 \le i \le l$, hence Using the Stirling's bound on $n!$, given by
\begin{eqnarray} \label{3:stir} n! \ge \sqrt{2 \pi n} (n/e)^n, \end{eqnarray}
we obtain
\[\left|m_{\MP}^{(l)}\right| \le l \frac{(y_bl^2)^{l}}{2 \pi l \left(l/e\right)^{2l}} < (y_be^2)^{l} \le (e^2Y_b)^{l}. \]
Now suppose that $y_b <1$. If $i_0=0$ or $1$, then the equality (\ref{3:bmp}) can be easily verified. If $i_0 \ge 2$, then $i_0 >\sqrt{y} l-1 \ge \sqrt{y} l/2$. Using the above Stirling's bound on $n!$ again, we obtain
\[\left|b_{\MP}^{(l)}\right|< l \frac{(yl^2)^{i_0}}{4 \pi \left(\sqrt{y}l/2e\right)^{2i_0}} < l(4e^2)^{\sqrt{y_b}l} \le (8e^2)^{l}. \]
This completes the proof of Lemma \ref{3:lem1}. \quad $\square$

Now from Lemma \ref{3:lem1} and \cite[Page 92]{bai}, we have
\begin{eqnarray} \label{3:bai}
\left|\sum_{i=1}^{l}y_a^{l-i+1} \sum_{\substack{k_1+k_2+\cdots+k_s=l-i+1\\
k_1+2k_2+\cdots+ik_i=l}}\frac{l!}{i!} \prod_{j=1}^i \frac{m_{\MP}^{(j)}(y_b)^{k_j}}{k_j!}\right| \le y_a(8e^2)^lY_b^l (1+\sqrt{y_a})^{2l} \le (32e^2)^lY_a^{l+1}Y_b^l.
\end{eqnarray}

\subsection{Proof of Theorem \ref{1:thm}} Using notation from Section \ref{sec-moment}, for each $\vec{s} \in \Omega_{N_a}^{(a)} \times \Omega_{N_b}^{(b)}$, let $\lambda_1(\vec{s}), \ldots, \lambda_{N_a}(\vec{s})$ be the eigenvalues of the matrix $\frac{1}{N_aN_b}\vec{A}^*_{s^{(a)}}\vec{A}_{s^{(a)}}\vec{B}^*_{s^{(a)}}\vec{B}_{s^{(a)}}$. The characteristic function we consider is
\[\phi(t):=\frac{1}{N_a} \sum_{k=1}^{N_a} \E\left(\exp\left(it (\lambda_k(\vec{s})-y_a)\right),\pp\right).\]
Let $\overline{M}_{\boxplus^2\MP}(x;y_a,y_b)$ be the distribution of the free multiplicative convolution of the Marchenko-Pastur densities $\mu_{\MP}(x;y_a)$ and $\mu_{\MP}(x;y_b)$ and let
\[M_{\boxplus^2\MP}(x;y_a,y_b):=(1-y_a)+y_a\overline{M}_{\boxplus^2\MP}(x;y_a,y_b).\]
Let $\vec{x}$ be a random variable with distribution $M_{\boxplus^2\MP}(x;y_a,y_b)$. It is known that
\[\E(\vec{x}^l)=\sum_{i=1}^{l}y_a^{l-i+1} \sum_{\substack{k_1+k_2+\cdots+k_s=l-i+1\\
k_1+2k_2+\cdots+ik_i=l}}\frac{l!}{i!} \prod_{j=1}^i \frac{m_{\MP}^{(j)}(y_b)^{k_j}}{k_j!}.\]
We shall consider
\[\gamma(t):=\E\left(\exp(it (\vec{x}-y_a) )\right). \]
Define for each $l$
\begin{eqnarray*} B_l=\frac{1}{N_a}\sum_{k=1}^{N_a} \E \left((\lambda_k(\vec{s})-1)^l, \pp\right). \end{eqnarray*}
Expanding the $l$-th power we find that
\begin{eqnarray} \label{3:bl} B_l=\sum_{t=0}^l (-1)^{l-t} \binom{l}{t} \E \left(A_t(\vec{s}), \pp\right), \end{eqnarray}
where estimates on $\E \left(A_t(\vec{s}), \pp\right)$ is provided by Theorem \ref{2:moment}. Using the inequality
\[\left|\exp(it)-\sum_{l=0}^{r-1}\frac{(it)^l}{l!}\right| \le \frac{|t|^r}{r!}, \]
and choosing the integer $r \ge 4$ to be even, we find that
\begin{eqnarray} \label{3:phi} \left|\phi(t)-\sum_{l=0}^{r-1}\frac{(it)^l B_l}{l!}\right| \le \frac{t^r B_r}{r!},\end{eqnarray}
and
\begin{eqnarray} \label{3:gamma} \left|\gamma(t)-\sum_{l=0}^{r-1}\frac{(it)^l \E((\vec{x}-y_a)^l)}{l!}\right| \le \frac{t^r \, \E((\vec{x}-y_a)^r)}{r!} .\end{eqnarray}
We note that $B_l=\E((\vec{x}-y_a)^l)$ for $l=0$ and $1$. For $l \ge 2$, using the expression (\ref{3:bl}) and Theorem \ref{2:moment}, given that $d^{\bot} \ge 5$, we find
\begin{eqnarray} \label{3:bll} \left|B_l-\E((\vec{x}-y_a)^l)\right| \le \sum_{t=2}^l \binom{l}{t} y_a^{l-t} \frac{t^{6t}Y_a(Y_aY_b)^t}{\min\{N_a,N_b\}} < \frac{(l^6+1)^l Y_a^{l+1}Y_b^l}{\min\{N_a,N_b\}}. \end{eqnarray}
As for $\E((\vec{x}-y_a)^l)$, from (\ref{3:bai}) we obtain
\begin{eqnarray} \label{3:x} \left|\E((\vec{x}-y_a)^l)\right| \le \sum_{t=0}^l \binom{l}{t}y_a^{l-t}(16e^2)^tY_a^{t+1}Y_b^t <(1+16e^2)^lY_a^{l+1}Y_b^l. \end{eqnarray}

In writing
\[|\phi(t)-\gamma(t)| \le \left|\phi(t)-\sum_{l=0}^{r-1}\frac{(it)^l B_l}{l!}\right|+\left|\gamma(t)-\sum_{l=0}^{r-1}\frac{(it)^l \E((\vec{x}-y_a)^l)}{l!}\right|+\left|\sum_{l=0}^{r-1}\frac{(it)^l
\left\{B_l-\E((\vec{x}-y_a)^l)\right\}}{l!}\right|, \]
and using the above estimates from (\ref{3:phi})--(\ref{3:x}) and Lemma \ref{3:prob}, we collect terms together and finally obtain
\begin{eqnarray*} \label{3:mcest} \left|M_{\cf}(x+y_a)-M_{\boxplus^2\MP}(x+y_a;y_a,y_b)\right| < \frac{(16e^2+1)^r Y_a^{r+1}Y_b^rT^r}{r! r}+\frac{Y_a^r Y_b^r r^{7r}T^r}{\min\{N_a,N_b\} r! r} + \frac{24 \cdot c(y_a,c_b) }{\pi T}, \end{eqnarray*}
where we are content with the use of the constant $c(y_a,y_b)$ which is an upper bound of the absolute value of a derivative of the distribution $M_{\boxplus^2\MP}(x+y_a;y_a,y_b)$, depending on $y_a,y_b$ only. 

Finally, taking $r$ to be a positive even integer of size
\[ r \approx \frac{c(y_a,y_b) \log n}{\log \log n}, \mbox{ and } \,\, T =\frac{r}{2e(16e^2+1)Y_aY_b},\]
where $c(y_a,y_b)$ is an appropriate constant which may be different from each appearance, and using the Stirling's bound (\ref{3:stir}), we see that as $n$ (and consequently $r$) is sufficiently large, all the three terms on the right side of (\ref{3:mcest}) can be bounded
\[ \frac{c(y_a,y_b) \cdot \log \log n }{\log n},\]
for some appropriate constant $c(y_a,y_b)$. This completes the proof of Theorem \ref{1:thm}. \quad $\square$






\end{document}